\documentclass[9pt,twocolumn,twoside]{osajnl}
\journal{josab} 

\ifthenelse{\boolean{shortarticle}}{\colorlet{color2}{color2b}}{\colorlet{color2}{color2}} 

\RequirePackage{graphicx}
\usepackage{graphicx}
\usepackage{dcolumn}
\usepackage{epstopdf}
\usepackage{bm}
\usepackage{mathpazo}
\usepackage{array}
\usepackage{pbox}

\title{Generation of orbital angular momentum through twisted Variable retarders }

\author[1,2]{Dipan Sinha}
\author[2,*]{Dipti Banerjee}

\affil[1]{Department of Physics, University of Calcutta,92,A.P.C Road,Kolkata-700009, West Bengal,INDIA}
\affil[2]{Department of Physics, Vidyasagar College for Women, 39,Sankar Ghosh lane,Kolkata-700006, West Bengal,INDIA}

\affil[*]{Corresponding author (Regular Associate of ICTP,Trieste,Italy) : deepbancu@hotmail.com}

\dates{Compiled \today}

\ociscodes{(270.5585)Quantum information and processing,(350.1370 )Berry's phase , (230.5440)Polarization-selective devices}

\doi{\url{http://dx.doi.org/10.1364/ao.XX.XXXXXX}}
\begin{document}
\begin{abstract}
Polarized light passing through twisted special wave plates show entangled state of both SAM and OAM. The variable circular and linear retarders (VCR and VLR) are studied here from the view point of quantum processing of entangled states. Interestingly the geometric phase in association with twist dependent gain of OAM has been discussed here. As an application of VLR, the origin of gradual flattened shape of wave fronts through successive array of antennas (QHQ) has been explained through new physical approach of the concurrence of entangled states.
\end{abstract}

\maketitle
\thispagestyle{fancy}

\section{Introduction}
A light beam has well defined SAM that describe polarization.The phase fronts of light beams possess orbital angular momentum (OAM) along their direction of propagation \cite{paper1}.The OAM can be of further two types(i)an external one,that aries from the cross product of the total momentum (p) and the position vector (r) with respect to the origin of coordinates (ii) an internal OAM component that is associated with the helical structure of the wave front around the beam axis and with an optical vortex located at the beam axis. A particular choice could eliminate the external OAM.For a paraxial beam having an azimuthal phase dependence $exp(il\varphi)$, OAM Hilbert space eigenstates denoted by  $|l\rangle_0$,correspond to with integer $l$ ,where $\varphi$ is the azimuthal angle around the beam axis \cite{paper2}\cite{paper3}\cite{paper4}.

Photon SAM can be manipulated by polarizers and birefringent plates. Cylindrical lens converters and Dove prisms can be used for changing OAM slowly \cite{paper5}.With every value of OAM, there are two polarized photons for which there exists 2 to 1 correspondence between SAM and OAM sphere. It was thought that these two different angular momentum of light are largely independent and non-interacting. Recently they are seen to coupled through specific complex media where a change of SAM modifies the OAM and vice versa. Marrucci et al  proposed that in anisotropic inhomogeneous media the variation of SAM occurring from the mediums birefringence gives rise to the appearance of OAM,arising from medium's inhomoginity \cite{paper6}.

The OAM beams are generated by a kind of birefringent plate known as "q-plates" \cite{paper7} which have very fruitful applications in the classical and quantum regime. Agarwal also pointed out \cite{paper8} that that q plate is the device which produces entangled state of orbital degree of freedom(OAM) and spin degrees of freedom (SAM) resulting the involvement of degrees of freedom optical medium.The property of birefringence develops quantum phases in the optical material.

This appeared phases may be either dynamical or geometrical or a mixture of both. There are four different geometric phases (GP) \cite{paper9}. In the context of polarization optics Pancharatnam first studied \cite{paper10} the phase $\Omega/2$,where $\Omega$ is the solid angel enclosed by the path of the light ray on the Poincare sphere. The physical mechanism of different kind of GP's originate from spin or orbital angular momentum of polarized photon. van Enk \cite{paper11} pointed out that Pancharatnam phase in mode space is associated with spin angular momentum (SAM) transfer of light and optical medium. The GP for OAM of polarized photon has been studied theoretically by Padgget \cite{paper12} in Poincare sphere and experimentally by Galvez et al.\cite{paper13} in mode space

Bhandari studied the details of GP in various combination of optical material \cite{paper14}. Inspired by the works' of Bhandari on variable retarders in connection with geometric phase \cite{paper15}, we here at first like to study the variable circular retarder (VCR) and variable linear retarder (VLR) from the view point of quantum  entanglement. In addition we will study the application of VLR in antenna array in the context of quantum information processing.

\section{Matrix representation of birefringent media}

The polarization of the incident light if remains unaltered after going through an optical medium,
the state can be identified as the eigenvector $\vec{D}_i$ of the
optical component $M_i$ following the eigen value equation
\begin{equation}
M\vec{D}_i=d_i\vec{D}_i
\end{equation}
where $d_i$ is the corresponding eigenvalue of a particular polarization matrix
$M=\begin{pmatrix}
m_{1} & m_{2} \\
m_{3} & m_{4}
\end{pmatrix}$.\\
The optical properties such as birefringence and dichroism of a homogeneous medium varies with distance.The passage of light through an optical element such as birefringent, absorbing or dichroic plate would be to change both $d_x$ and $d_y$ so that the effect may be represented by $|\psi_f>=M|\psi_i>$.
For a non-absorbing plate, there is no change in the intensity and the polarization matrix $M$ is therefore unitary $detM=1$ which makes $|\psi_f|=|\psi_i|$.

The property of birefringence of this optical medium can be represented by the differential matrix $N$. At a particular position z of optical media, Jones \cite{paper16} showed that the spatial variation of the polarization matrix $M$ develops the $N$ matrix. For a particular wavelength through an infinitesimal distance within the optical element the propagation of the light vector $\varepsilon$ can be represented by
\begin{equation}
\frac{d\varepsilon}{dz}=\frac{dM}{dz}\varepsilon_0
=\frac{dM}{dz}M^{-1}\varepsilon=N\varepsilon
\end{equation}
where it is evident that $N$ is the operator that determines $dM/dz$ from $M$ as follows
\begin{equation}
N=\frac{dM}{dz}M^{-1}=
\begin{pmatrix}
n_1 & n_2 \\
n_3 & n_4
\end{pmatrix}
\end{equation}

Photon is a two state system and can exist in an arbitrary superposition $\alpha|0\rangle+\beta|1\rangle$ as a computational resource, where $\vert 0 \rangle=\begin{pmatrix}
1 \\
0
\end{pmatrix}$ and
$\vert 1 \rangle=\begin{pmatrix}
0 \\
1
\end{pmatrix}$.
These are the standard unit of quantum information. Using elementary qubits an arbitrary incident polarized photon can be written as \cite{paper17} \cite{paper18}.
\begin{equation}
\vert \psi \rangle = {e^{i\varphi} cos\theta/2} \vert 0 \rangle+sin\theta/2\vert 1 \rangle \approx e^{i\varphi} {\begin{pmatrix}
{Y^{0}_{1}} \\
{Y^{1}_{1}}
\end{pmatrix}}
\end{equation}
 which could construct the polarization matrix. Here $Y^{l}_{m}$ are the spherical harmonics.

 Following Berry \cite{paper19} this polarization matrix $(|\psi\rangle \langle\psi|-1/2)$ becomes

\begin{equation}
 M = \left( \begin {array}{cc} {{Y^{0}_{1}}{Y^{*0}_{1}}-{1/2}}&{{Y^{0}_{1}}{Y^{*1}_{1}}}\\ \noalign{\medskip}{{Y^{1}_{1}}{Y^{*0}_{1}}}&{{Y^{1}_{1}}{Y^{*1}_{1}}-{1/2}}\end {array}\right)
\end{equation}
\begin{equation}
M=\begin{cases}
 \frac{1}{2}\left( \begin {array}{cc} {Y^{0}_{1}}&{Y^{1}_{1}}\\ \noalign{\medskip}{Y^{-1}_{1}}&{-{Y^{0}_{1}}}\end {array}\right)\\\\
 \frac{1}{2}\left( \begin {array}{cc} {\cos\theta}&{\sin\theta e^{i\varphi}}\\ \noalign{\medskip}{\sin\theta e^{-i\varphi}}&{-\cos\theta}\end {array}
\right) \end{cases}
\end{equation}
This polarization matrix has been determined earlier \cite{paper20},and realized lying on the OAM sphere for $l=1$. For the conjugate state
\begin{equation}
\vert \tilde{\psi} \rangle = e^{-i\varphi}{\begin{pmatrix}
{Y^{0}_{1}} \\
{-Y^{-1}_{1}}
\end{pmatrix}}
\end{equation}

\begin{figure}[htbp]
\centering
\fbox{\includegraphics[width=7cm]{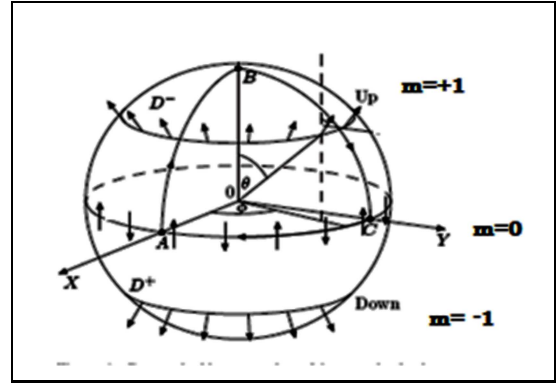}}
\caption{OAM corresponding $l=1$ with $m=+1,0,-1$.For upper hemisphere $m=+1$ and for lower hemisphere $m=-1$, the equatorial linerepresented by $m=0$.}
\end{figure}
the corresponding polarization matrix
\begin{equation}
\tilde{M}=\begin{cases}
 \frac{1}{2}\left( \begin {array}{cc} {-{Y^{0}_{1}}}&{Y^{-1}_{1}}\\ \noalign{\medskip}{Y^{1}_{1}}&{Y^{0}_{1}}\end {array}\right)\\\\
 \frac{1}{2}\left( \begin {array}{cc} {-\cos\theta}&{\sin\theta e^{i\varphi}}\\ \noalign{\medskip}{\sin\theta e^{-i\varphi}}&{\cos\theta}\end {array}
\right) \end{cases}
\end{equation}
We may construct the OAM sphere for $l=1$ with corresponding magnetic quantum no $m=1,0,-1$ . The polarization state $|\psi\rangle$ eq.(3) and polarisation matrix M from eq.(5) exist on the upper hemisphere for $m=1$ and $|\tilde{\psi}\rangle$  and polarization $\tilde{M}$ from eq.(7) on the lower hemisphere for $m=-1$. Here we consider the polarization state on the upper hemisphere only. For higher OAM states l=2,3.., further study is needed to evaluate the polarization matrix for a particular OAM from the respective product harmonics $Y^{m}_{l}$.

Considering $z=(cos\theta$),the thickness of the optical medium, the $N$ matrix can be obtained \cite{paper17} from $M$ of eq.(6), where $\theta$ is the angular variable of light after refraction.
\begin{equation}
 N = \eta {\left( \begin {array}{cc} 0&{-e^{i\varphi}}\\ \noalign{\medskip}{e^{-i\varphi}}&0\end {array}\right)}
\end{equation}

The internal birefringence $\eta$ visualized through eigenvalues $±i\eta$ have dependence on $\theta$.As the birefringent crystal is twisted about the direction of transmission, the $N$ matrices are transformed to $N(\vartheta)$ by rotation matrix $S(\vartheta)$ upon rotation.
\begin{equation}
N(\vartheta)=S(\vartheta)N S(-\vartheta)
\end{equation}
This angle of twist has dependence on crystal parameter $N(\vartheta)$
\begin{equation}
N(\vartheta)= \eta \cos\varphi \left( \begin {array}{cc} 0&-1\\ \noalign{\medskip}1&0\end {array} \right) -i \eta \sin\varphi \left( \begin {array}{cc} {-\sin  2\,\vartheta } &{\cos 2\,\vartheta } \\ \noalign{\medskip}{\cos  2\, \vartheta }&{\sin  2\,\vartheta } \end {array} \right)
\end{equation}
 We like to consider the incidence of linearly polarized light on the above twisted matrix.
The horizontal and vertical component of linearly polarized light can be written in terms of LCP and RCP in the following manner
\begin{eqnarray}
|h\rangle &=& \frac{1}{\sqrt{2}}(|L\rangle+|R\rangle)\nonumber\mbox and\\
|v\rangle &=& \frac{1}{\sqrt{2}}(|L\rangle-|R\rangle)
\end{eqnarray}
 where using elementary qubits the circular and linear polarisation states of photon
 $\vert L \rangle=\frac{1}{\sqrt{2}}\left(\vert0\rangle+i\vert1\rangle\right)$ and $\vert R \rangle=\frac{1}{\sqrt{2}}\left(\vert0\rangle-i\vert1\rangle\right)$
The passage of the horizontal component of linearly polarized light through this twisted birefringent plate results.
\begin{equation}
N(\vartheta)|h\rangle=-i\eta cos\varphi |v\rangle + \frac{\eta}{\sqrt{2}}  sin\varphi( |R\rangle e^{i2\vartheta}-|L\rangle e^{-i2\vartheta})
\end{equation}
The developed emergent wave has combined polarization of OAM and SAM degree of freedom in association with vertical direction of polarization.It has been realized here that this twisted birefringent media shows similar behavior of q-plates. Agarwal \cite{paper8} pointed out that q-plate helps to couple the spin and orbital angular momentum. He equated q-plate with the twisted birefringent plate such as Half wave plate(HWP) which is in the xy plane having optic axis making an angle $\vartheta$ with x axis. For the standard half wave plates $\vartheta$ is a constant. This angle $\vartheta(\theta)=q\theta+\vartheta_0 $ will be dependent on some coordinate when this half wave-plate is made of nematic liquid crystal. He showed that incident RCP changes to LCP acquiring the orbital angular momentum with phase
$e^{2i(q\theta+\vartheta_0)}$.
This idea of q-plate is is also reflected in \cite{paper4}
\begin{equation}
\bar{QP}|h\rangle_{\pi}|m\rangle_{0}=[|L\rangle_{\pi}|m-2q\rangle_{0}+|R\rangle_{\pi}|m+2q\rangle_{0}]
\end{equation}
In the next section we will study the quantum information through variable circular retarders made of special half wave plates.

\section{variable retarders}
\subsection{Variable circular  retarder(VCR)}
An untwisted $N$ matrix (eq 9)for $\varphi=\pi/2$ is represented by a half wave- plate($H$). The passage of LCP and RCP through half wave plate though changes only SAM, the nontrivial twist of special Half wave plate by $\vartheta$ changes $H$ to $H(\vartheta)$ with the appearance of twist dependent OAM in the incident circularly polarized light (LCP or RCP).
\begin{equation}
H(\vartheta)|L\rangle=\eta|R\rangle e^{2i\vartheta}
\end{equation}
similarly
\begin{equation}
H(\vartheta)|R\rangle=\eta|L\rangle e^{-2i\vartheta}
\end{equation}

This confirms that twisted half wave-plate change the SAM $|L\rangle$ to $|R\rangle$ and OAM of the incident polarized light.

Variable circular retarder (VCR) introduces variable phase between RHC and LHC by rotating an arbitrary state on the Poincare sphere about the polar axis\cite{paper15}.If the polarized light from first wave plate $H$ is made to pass through another wave plate $H(\vartheta)$ the combination is as similar as VCR.
\begin{equation}
HH(\vartheta)|L\rangle=-\eta^{2}|L\rangle e^{2i\vartheta}
\end{equation}
and
\begin{equation}
HH(\vartheta)|R\rangle=-\eta^{2}|R\rangle e^{-2i\vartheta}
\end{equation}
The polarized light passing through this VCR traces a closed area on a Poincare sphere for which a geometrical phase of Pancharatnam \cite{paper9} becomes
\begin{equation}
\gamma^{L}_{c}= \langle L \vert HH(\vartheta) \vert L \rangle = -{\eta}^{2} e^{i2\vartheta}
\end{equation}
and
\begin{equation}
\gamma^{R}_{c}= \langle R \vert HH(\vartheta)\vert R \rangle = -{\eta}^{2} e^{-i2\vartheta}
\end{equation}
If we apply arbitrary twist of angle $\alpha$ between the two H plates, LCP acquire similar OAM dependent geometric phase in terms of $2\alpha$ without dependence of SAM.
\begin{equation}
 H(\omega+\alpha)H(\omega)|L\rangle=-i\eta^2 |L\rangle e^{2i(\omega+\alpha)}
\end{equation}
Even number of half wave plates develops non-zero OAM dependent geometric phase by the change of OAM with no change in SAM. Both the change of SAM and OAM is visible if we apply LCP through odd number of times the half wave-plates.
\begin{equation}
 H(\omega+\alpha+\beta)H(\omega+\alpha)H(\omega)|L\rangle = -i\eta^2|R\rangle e^{2i(\omega+\beta)}
\end{equation}
Hence,in both the cases gain of orbital angular momentum is observed \cite{paper21}.
The above equation is quite similar to the followings  equation for q-plate \cite{paper7} \cite{paper8}.
\begin{equation}
\bar{QP}|L\rangle_{\pi}|m\rangle_{0}= |R\rangle_{\pi}|m+2q\rangle_{0}
\end{equation}

Where here the topological charge q is equivalent to the angle of twist of the wave plates that is the source of anisotropy to the system.

The propagation of polarized light through variable linear retarder will be our next interest whose application in antenna array will be studied from the view point of quantum entanglement.

\subsection{Variable linear  retarder(VLR)}
The gadget that introduces variable phase between the linearly polarized light by rotating an arbitrary state on the Poincare sphere about the equatorial line is called Variable linear retarder (VLR).According to Bhandari \cite{paper15}, VLR can be represented by sandwiching one HWP at an angle $\alpha$ in between two quarter wave plates (QWP's) at identical angle $\vartheta$ in the configuration $Q(\vartheta)H(\alpha)Q(\vartheta)$  . Using eq.(11)  we calculate the required QHQ gadget for OAM l=1.
\begin{eqnarray}
\noindent
\scriptstyle\hspace*{-0.5cm}
Q(\vartheta)H(\alpha)Q(\vartheta)=\frac{-i{\eta}^{3}}{2} [[S(2\alpha)-S(4\vartheta+2\alpha) ]\sigma_{x}-[2S(2\vartheta+2\alpha) \sigma_{y}]]
\noindent
\end{eqnarray}
where $S(2\vartheta)$ is the rotation matrix and $\sigma_{x}$ and $\sigma_{y}'s$ are the Pauli's spin matrices.
In the context of polarization optics Pancharatnam \cite{paper22} pointed out that if the first and the third QWP's has arbitrary retardation $2\delta_{1}$  and middle HWP has retardation $2\delta_{2}$  the combination will act as a linear retarder. Inspired by  the works of Bhandari \cite{paper14} \cite{paper23} also, we here studied the quantum information processing through the VLR by the passage of linearly horizontal polarized state $|h\rangle$ whose OAM is initially zero.

\begin{equation}
\ Q(\vartheta)H(\alpha)Q(\vartheta)|h\rangle=\begin{cases} \frac{-i\eta^{3}}{2}
[(i(\vert R \rangle e^{i2\alpha}-\vert L \rangle  e^{-i2\alpha})\\
-i(\vert R \rangle e^{i(4\vartheta+2\alpha)}-\vert L \rangle e^{(-i(4\vartheta+2\alpha)})\\
-2(\vert L \rangle e^{(-i(2\vartheta+2\alpha))}+\vert R \rangle e^{i(2\vartheta+2\alpha)})]
\end{cases}
\end{equation}
The above equation visualizes the emergent ray as the combination of three entangled states of SAM and OAM of linearly polarized light. The power of OAM in the entanglement states increases with the orientation of twisted HWP which is sandwiched between two QWP's at same orientation. That means the turning of external OAM can be controlled by the angle of twist which is dependent on medium's parameter. Also the states are similar (except the number of states) to the final state of q-plates\cite{paper7} \cite{paper8} where OAM generated when a polarized light passes through twisted birefringent media.
 Eq.(22) can be represented by combination of three singlet state $|\psi_{1}\rangle, |\psi_{2}\rangle and |\psi_{3}\rangle$
\begin{equation}
|\Pi\rangle= \frac{-i\eta^{3}}{2}[|\psi_1\rangle -|\psi_2\rangle -2i|\psi_3\rangle ]
\end{equation}
where $|\psi_{1}\rangle=(i(|R\rangle e^{i2\alpha}-|L\rangle {e^-i2\alpha}$,$|\psi_{2}\rangle =i(|R\rangle e^{i(4\vartheta+2\alpha)}-|L\rangle e^{(-i(4\vartheta+2\alpha)})$ and $|\psi_{3}\rangle=(|L\rangle e^{(-i(2\vartheta+2\alpha))}+|R\rangle e^{i(2\vartheta+2\alpha)})$
the phase difference between incident and emergent light
\begin{equation}
\ \langle h|\Pi\rangle= \gamma_{\alpha}=2\eta^3 [sin2\alpha-sin(4\vartheta+2\alpha)-2isin(2\vartheta+2\alpha)]
\end{equation}

\begin{figure}[htbp]
\centering
\fbox{\includegraphics[width=6cm]{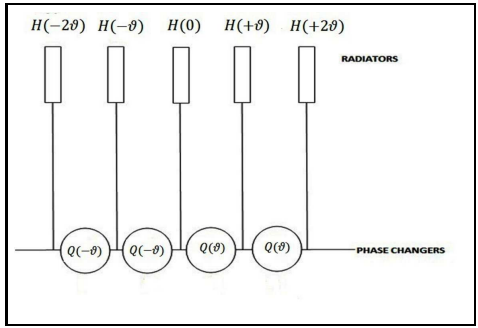}}
\caption{ Antenna equipped with QHQ phase shifter whose middle HWP's are integer multiple of the angle of twist(adapted from $fig 1$ of reference $24$).}
\label{fig:2}
\end{figure}

The gadget QHQ thus produces a linearly increasing phase shift as a function of the orientation of the Half wave plate and as the phase shifter  becomes integer multiple of $\pm\vartheta$, the geometrical phase increases gradually.
 This principle was first used by Fox \cite{paper24} to introduce a linearly increasing phase difference between successive elements of an array of radar antennas to enable to radiate in a chosen direction (Fig.3).In optics,where one overlooks the wavelength reliance of the refractive index it will be hard to see the sign inversion due to the restricted extent in wavelength of the limited range. However, in radio physics, applications of this kind was considered by Fox, where one could apply such stage shifters in varieties of radio receiving wires, for case to guide the light emission cluster of radio telescopes, the above impacts will produce interesting results. In optics Bhandari \cite{paper23} uses the same law of variation of retardation with wavelength (Fig.3) in case of antenna equipped with QHQ phase shifter.
\begin{align}
Q(\vartheta)H(\vartheta)Q(\vartheta)=\frac{-i\eta^{3}}{2}[[S(2\vartheta)-S(6\vartheta) ] \sigma_x\\\notag
& \hspace{-0.8cm}-[2S(4\vartheta) ] \sigma_y ]
\end{align}
and
\begin{align}
Q(-\vartheta)H(-\vartheta)Q(-\vartheta)=\frac{-i\eta^{3}}{2}[[S(-2\vartheta)-S(-6\vartheta) ]\sigma_x\\\notag
& \hspace{-3cm}-[2S(-4\vartheta) ]\sigma_y ]
\end{align}
This pair of equation shows that $Q(\vartheta)H(\theta)Q(\vartheta)$ and $Q(-\vartheta)H(-\vartheta)Q(-\vartheta)$ is the inverse of each other in rotation. Corresponding geometrical phase are
\begin{equation}
\gamma_{\vartheta}= 2\eta^{3}[sin2\vartheta-sin6\vartheta-2isin4\vartheta ]
\end{equation}
and
\begin{equation}
\gamma_{-\vartheta} = 2\eta^{3}[sin2\vartheta-sin6\vartheta-2isin4\vartheta ]
\end{equation}
Now applying successive VLR whose middle HWP's are integer multiple of the angle of twist to the horizontal component of linearly polarized light, we found that angle of twist $(\vartheta)$ of the medium creating a twist dependent external  OAM (i.e the additional helicity in the emergent wave front)
\begin{equation}
Q(\vartheta)H(\vartheta)Q(\vartheta)|h\rangle=\begin{cases}\frac{-i\eta^{3}}{2}[i({\vert R \rangle} e^{i2\vartheta}-\vert L \rangle e^{-i2\vartheta})\\
-i(\vert R \rangle e^{i6\vartheta}-\vert L \rangle e^{-i6\vartheta})\\
-2(\vert L \rangle e^{-i4\vartheta}+\vert R \rangle e^{i4\vartheta})]\\\\
=|h'\rangle \end{cases}
\end{equation}

\begin{eqnarray}
\noindent
\scriptstyle\hspace*{-0.5cm} Q(\vartheta)H(2\vartheta)Q(\vartheta)| h'\rangle=\begin{cases}\noindent
\scriptstyle \frac{-\eta^{6}}{4}[2(|L\rangle e^{-i2\vartheta}+| R \rangle e^{i2\vartheta})\\ \noindent
\scriptstyle
-(| L \rangle e^{i2\vartheta}+| R \rangle e^{-i2\vartheta})-(| L \rangle e^{-i6\vartheta}+| R  \rangle e^{i6\vartheta})\\ \noindent
\scriptstyle
+2(| L \rangle-| R \rangle)-2(|R\rangle e^{i4\vartheta}-|L\rangle e^{-i4\vartheta})\\ \noindent
\scriptstyle
-2i(\vert L \rangle e^{-i8 \vartheta}+\vert R \rangle e^{i8\vartheta})-2i(| L \rangle e^{-i 12 \vartheta}+\vert R \rangle e^{i 12 \vartheta})\\ \noindent
\scriptstyle
-2(\vert L \rangle e^{-i 10 \vartheta}+\vert R \rangle e^{i 10 \vartheta})]
\noindent\\\\
=\vert h'' \rangle \end{cases}
\end{eqnarray}
\begin{figure}[htbp]
\centering
\fbox{\includegraphics[width=7cm]{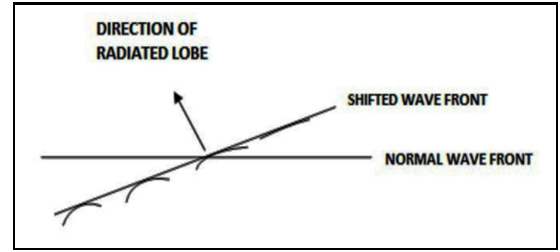}}
\caption{Wave front gradually flattened as it passes through such successive array of antennas(adapted from $fig 1$ of reference $24$).}
\label{fig:3}
\end{figure}

Each bracketed {} states in eq.(31)and eq.(32) are Bell's state $|e_{i}\rangle=\frac{1}{2}[|\uparrow\downarrow\rangle \pm |\downarrow\uparrow\rangle]$  developed from SAM and OAM degree of freedom of polarized light passing through VLR.
 The power of entanglement of SAM and external OAM (twist dependent) is increased gradually from the state $|h'>$ to $|h''>$ through the corresponding angle of twist (i.e $\vartheta$) of the $H$-plate in $QHQ$. As the state $\psi=\sum_{i}\alpha_{i}|e_{i}\rangle$ is in the particular basis, its measure of entanglement can be expressed in terms of concurrence $C=|\sum_{i}\alpha_{i}^{2}|$ \cite{paper25}. This quantity $C$ achieves a maximum and minimum value and has dependence on internal birefringence(i.e $\theta$ or most specifically the angle of incident) and the angle of twist($\vartheta$) of the optical medium.
The respective concurrence for $h'$ and $h''$ are
\begin{equation}
C(h') = \frac{\eta^{6}}{4}[3-cos4\vartheta]
\end{equation}

\begin{align}
C(h'') = \frac{\eta^{12}}{16}[[5-3cos8\vartheta+2cos12\vartheta+2cos4\vartheta] \\\notag
& \hspace{-5cm} +i[4sin10\vartheta+8sin2\vartheta+4sin6\vartheta]
\end{align}
 In order to measure concurrence it is required to consider $2\vartheta=n\pi$ so that all the imaginary terms after being vanished, the final C is real. This further implies that the antenna will be more effective if the two Q's in QHQ are kept at perpendicular to the orientation of middle HWP.
 Variation of concurrence with $\vartheta$ are plotted below.Both the graphs (Fig 4 and Fig 5 ) obtained from the respective eq(31) and eq(32) visualize that the concurrence is depending on the angle of twist $(\vartheta)$ and the internal birefringence $\eta$ which is dependent on $\theta$, the angle of incidence.
 \begin{figure}[htbp]
\centering
\fbox{\includegraphics[width=6cm]{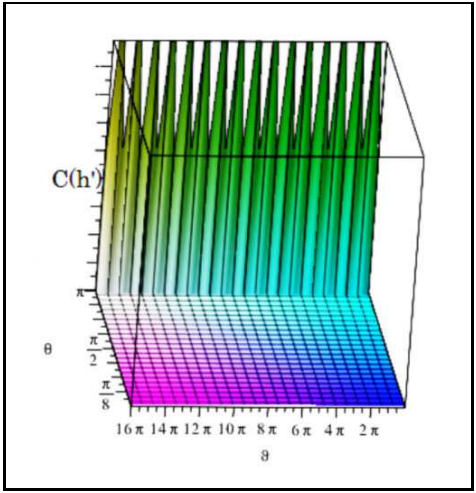}}
\caption{C($h^{\prime}$)[the concurrence when light passes through the first $Q(\vartheta)H(\vartheta)Q(\vartheta)$ of the antenna] vs $\vartheta$ [the angle of twist] and $\theta$ [the internal birefringence] }.
\label{fig:4}
\end{figure}
It is also visualized from both the graphs that the value of concurrence for $C(h'')$ is less than $C(h')$. The gradual flattening of wave-front as pointed out by Fox and Bhandari  \cite{paper14}\cite{paper23} supports our findings where the minimum value of concurrence decreases from $C(h')$ to $C(h'')$. From the view point of quantum entanglement we sketch here the application of antenna array equipped with VLR (QHQ) where the the orientation of HWP is $n\pi$ and that of Q's are $n\pi/2$. We can conclude that the gradual minimum value of concurrence assures the gradual flattening of the wave-front representing the superposition of entangled states produced through the transmission of circularly polarized light through successive variable linear retarder (VLR). This confirms the wave front gradually flattened as it passes through such successive array of antennas represented by gadget QHQ  where middle plate suffer successive rotation $\pm\vartheta$, from the nearest one as seen in (Fig.3).
\begin{figure}[htbp]
\centering
\fbox{\includegraphics[width=6cm]{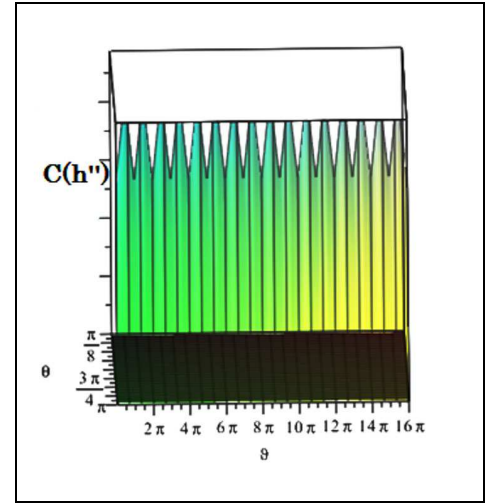}}
\caption{C($h^{\prime\prime}$)[the concurrence when light passes through the second $Q(\vartheta)H(2\vartheta)Q(\vartheta)$ of the antenna] vs  $\vartheta$[angle of twist] and $\theta$ [the internal birefringence] }.
\label{fig:5}
\end{figure}
\pagebreak
\section{Conclusion}
 We have considered here the polarization state by spherical harmonics ${Y_m}^l$ where $m$ varies between $l,0,-l$ to construct the polarization matrix $M$ on OAM sphere for $l=1$. Consequently the differential matrix $N$ representing birefringent medium has been evaluated which help to identify the twisted Quarter and Half wave plates. The proper combination of $Q$ and $H$ plates initiate to study here the variable retarders VCR and VLR. Obviously these retarders are their origin in OAM l=1. As the circularly polarized light is made to incident on the variable retarders made of non-trivial wave-plates where the angle of twist depends on medium's parameter, entangled state of SAM and external OAM (dependent of angle of twist) are observed. We found that VCR and VLR behave like "q"-plates by generating OAM beams. The strength of OAM in the entangled state increases by the corresponding change of the orientation of the HWP in VLR represented by $Q(\theta)H(\vartheta)Q(\theta)$.
 The physical origin of flattened shape of wave fronts has been realized by the real and diminishing values of concurrence of the successive entangled states through QHQ antenna when only the orientation of $H(2\vartheta)$ is $n\pi$ and perpendicular to the two $Q(\vartheta)$ at its both side.
 From the view point of quantum entanglement in the array of antenna (QHQ) made of nontrivial wave-plates $(Q,H)$, our present study
 shall initiate a new direction of research with q-plates.

\end{document}